\documentclass[11pt]{article}
\usepackage{amsmath}
\usepackage{amssymb}
\usepackage{amsthm}
\usepackage{graphicx}
\usepackage[margin=1in]{geometry}
\usepackage{setspace}
\usepackage{caption}
\usepackage{booktabs}
\usepackage{multirow}
\usepackage{array}
\newtheorem{theorem}{Theorem}
\newtheorem{lemma}{Lemma}

\usepackage{hyperref}
\hypersetup{pdftex,colorlinks=true,allcolors=blue}
\usepackage{hypcap}

\providecommand{\keywords}[1]{\text{Keywords:} #1}

\title{Matching Estimators for Causal Effects of Multiple Treatments}
\author{Anthony D. Scotina\footnote{Department of Mathematics and Statistics, Simmons University, Boston, MA 02115}, \ Francesca L. Beaudoin\footnote{Department of Health Services, Policy, and Practice, Brown University, Providence, RI 02912}\footnote{Department of Emergency Medicine, Brown University, Providence, RI 02912}, and Roee Gutman\footnote{Department of Biostatistics, Brown University, Providence, RI 02912}}
\date{}

\begin{document}

\maketitle

\begin{abstract}
Matching estimators for average treatment effects are widely used in the binary treatment setting, in which missing potential outcomes are imputed as the average of observed outcomes of all matches for each unit. With more than two treatment groups, however, estimation using matching requires additional techniques. In this paper, we propose a nearest-neighbors matching estimator for use with multiple, nominal treatments, and use simulations to show that this method is precise and has coverage levels that are close to nominal. In addition, we implement the proposed inference methods to examine the effects of different medication regimens on long-term pain for patients experiencing motor vehicle collision. \\

\noindent
\end{abstract}

\section{Introduction}

\subsection{Overview}

Many real-world applications of statistics involve comparison of multiple interventions or treatments. Randomized experiments are the preferred scientific approach to obtaining an unbiased comparison of two or more interventions. Multi-arm randomized experiments have been proposed as an efficient experimental design to identify among multiple active interventions those that are significantly better than a control treatment \cite{royston-03, jaki-15}. In other cases, multi-arm randomized experiments are used to identify the optimal active interventions \cite{dunnett-84}. When the design and implementation of randomized experiments is untenable because of financial, logistical, or ethical considerations, non-randomized observational studies can be used to compare the effectiveness of different interventions.

With continuous outcomes, a common practice is to conduct a ``global test" to compare whether one of the means is different. Only if the global test is rejected, additional analyses to identify specific differences between the treatments are conducted \cite{dagostino-91, rosner-05, cabral-08}. Various estimands were proposed to evaluate differences among multiple treatments in randomized experiments. Common estimands include all possible pairwise differences in means, all differences in means between the treatment arms and the best treatment, all differences in means between active treatments and a control, and all differences in means between the treatment arms and the overall mean \cite{rao-09}. In the context of randomized experiments, many statistical procedures were developed to provide interval estimates and hypothesis tests for these estimands (see \cite{rao-09} for references). These estimands are important in comparative effectiveness research, because it is often insufficient to only examine a single hypothesis of whether at least one of the treatments is different on average from the other treatments. 

Non-randomized observational studies have been proposed as possible designs to compare the effectiveness of multiple interventions for timely and urgent public health problems. For example, each year nearly 4 million Americans are evaluated in emergency departments (ED) after experiencing a motor vehicle collision (MVC) \cite{niska-10}. At the time of discharge from an ED, patients are typically prescribed an opioid analgesic and/or a nonsteroidal antiinflammatory drug (NSAID) \cite{beaudoin-17}. Prescribing of opioid analgesics has come under great scrutiny in the face of an ongoing epidemic of opioid addiction. Thus, comparison of the effectiveness of multiple pain medications regimens  on chronic pain is a pressing public health issue.  A similar type of public health issue requiring pairwise comparisons was necessary to study the cost-effectiveness of case management for patients with dementia \cite{vroomen-16}, as well as to compare changes in coronary atheroma volume in patients receiving high-intensity therapy, low-intensity statin therapy and no-statin therapy \cite{puri-15}.

Because individuals were not randomized to interventions in observational studies, individuals receiving one intervention may differ from those receiving another with respect to baseline covariates. Matched sampling has been proposed to balance units on pre-intervention characteristics to replicate the balance that would have occurred in randomized experiments. The theoretical basis of matching to remove covariate bias was developed in the 1970s with papers by Cochran and Rubin \cite{cochran-73} and Rubin \cite{rubin-73a,rubin-73b}, for settings with one covariate and two interventions. Matching procedures are comprised of three main components: a distance measure between two units, the matching algorithm, and the inference procedure for the matched cohorts. In this paper we provide inference methods for matching methods when comparing multiple treatments in observational studies. We apply the proposed methods to examine different pain medications regimens on long-term pain after motor vehicle collisions.

\subsection{Distance measures}
\label{subsec:distance}

When only a single covariate influences the assignment to treatment, it is often easy to identify similar units. This task is more complicated as the number of covariates increases. With multiple covariates and two treatment groups, the propensity score was proposed as a distance measure to identify similar units \cite{rosenbaum-83}.

The generalized propensity score \cite{imai-04} was proposed as a design tool to reduce the dimension of the covariate space with multiple covariates and multiple nominal interventions. In contrast to the propensity score, with $Z>2$ interventions, the generalized propensity score is a vector of $Z-1$ dimensions, which complicates the task of identifying similar units. Generally, both scores are unknown and are estimated from the observed data (see \cite{stuart-10} and \cite{lopez-15} for a review of estimation methods). One possible estimation model for the generalized propensity score is the multinomial logistic regression. 

Other possible distance measures include the Euclidean or Mahalanobis distance on the original covariates or on a function of the known or estimated generalized propensity scores. Once a distance measure is defined, matching algorithms can be used to identify similar units. 

\subsection{Matching algorithms for multiple treatments}

Lechner \cite{lechner-01, lechner-02} estimated the average treatment effects (ATE) for multiple treatments using a series of binary comparisons. For each pair of treatments the method matches on the estimated propensity score and separately estimates the average treatment effect for units receiving either of these two treatments.

Common referent matching is a matching method for three treatments that creates sets with one individual from each treatment group \cite{rassen-11}. The treatment group with the smallest sample size (say, group 1) is used as the reference group, and the propensity scores for treatment groups 1 and 2, and 1 and 3 are estimated separately using either logistic or probit regression models. Using 1:1 matching, pairs of units receiving treatments 1 or 2 are matched using the estimated propensity score, and similarly for units receiving treatments 1 or 3. Only units receiving treatment 1 which were matched to a unit receiving treatment 2 and to a unit receiving treatment 3, along with their associated matches constitute the sample used for analyses.

An issue with a series of binary comparisons and common referent matching is that treatment effect estimates only generalize to a subset of the population, rather than to the population of units eligible to receive all of the available treatments. This may result in non-transitive treatment effects, and prevents researchers from identifying the best treatment \cite{lopez-15}. 

Vector matching is a greedy algorithm that uses $k$-means clustering and 1:1 nearest neighbor matching to ensure that units matched on one component of the generalized propensity score vector are well matched on all of the other components \cite{lopez-15}. Vector matching relies on matching with replacement and had the lowest covariate bias in matched sets, compared to other bias reduction methods \cite{lopez-15}.

One issue with vector matching is its reliance on greedy matching, which may not be the most optimal procedure to reducing covariate bias among all treatments. With binary treatment, algorithms like full matching \cite{rosenbaum-89}, which relies on network flow theory, and mixed integer programming \cite{zubizarreta-12} were proposed to optimally match units such that the difference in the covariates' distributions between the two treatment groups is minimized while retaining most of the units. Optimally matching for multiple treatments, also known as $k$-dimensional matching, was shown to be a NP-hard problem \cite{karp-72}. The computational complexity of such methods may make them impractical in problems with many units and multiple treatments. 

\subsection{Inference procedures} 

For two treatments, the statistical literature include several procedures for point and interval estimates for the ATE with matched units. Generally, point estimates from matched units are obtained using similar procedures that are applied to entire datasets \cite{stuart-10}. Interval estimation has been debated in the literature. Randomization based standard errors have been shown to underestimate the true standard error, resulting in statistically invalid interval estimates \cite{abadie-06, hill-06, gutman-15}. For 1:1 matching with replacement, Hill and Reiter \cite{hill-06} proposed to use the Hodges-Lehmann aligned rank test when the treatment effect is additive, and when the treatment effect is not additive, they proposed a non-parameteric bootstrap algorithm for standard error estimates. When estimating the ATE, Abadie and Imbens \cite{abadie-08} showed that for matching with replacement the bootstrap method may be invalid in certain situations, and suggested using the formula derived in \cite{abadie-06} for estimating the standard error. For matching without replacement, Austin and Small \cite{austin-14} proposed a non-parametric bootstrap procedure for estimating the ATE after propensity score matching. Using simulations, they showed that estimates of the standard error using this procedure were close to the empirical standard deviation of the sampling distribution of the estimated effects. For multiple treatments, Rassen et al. \cite{rassen-11} relied on sampling variance estimates that ignore the variability in the matching procedure. As in the binary case, this estimate may underestimate the standard error. 

Non-matching estimation methods for binary and multiple treatments include linear regression adjustments, inverse probability weighting, and doubly robust methods \cite{mccaffrey-13}. For binary treatment, adjusting for covariate imbalances using linear regression adjustment has been found to be generally biased for estimating ATEs, and are only approximately unbiased when the two response surfaces are nearly linear and parallel \cite{rubin-73b}. Gutman and Rubin \cite{gutman-15, gutman-15a} found that both inverse probability weighting and doubly robust estimation resulted in generally valid procedure for binary treatment, though each of these methods is susceptible to extreme weights which can yield erratic causal estimates. This phenomenon is exacerbated with an increasing number of treatments and covariates that are not normally distributed \cite{lopez-15}. McCaffrey et al. \cite{mccaffrey-13} proposed to apply a sandwich estimator in combination with generalized boosted models to estimate the generalized propensity scores to obtain individual point and interval estimates for pairwise ATEs.

Yang et al. \cite{yang-16} proposed matching and subclassification estimates for estimating pairwise average treatment effects with multiple treatments. However, they did not provide an overall global test for the any difference between the outcomes, and they only describe an estimation procedure for the average differences over the entire population.

Abadie and Imbens provided an empirical formula for variance estimation of matching estimators for binary treatment \cite{abadie-06} and proposed a bias-corrected matching estimator that yielded consistent point estimates when there was more than one continuous covariate \cite{abadie-11}. We extend the work of Abadie and Imbens \cite{abadie-06, abadie-11} by deriving super-population point and interval estimates for the vector of the pairwise average treatment effects when comparing more than two treatments and using matching with replacement procedures. Our procedure enables to perform a global hypothesis test as well as derivation of individual point and interval estimates for all pairwise ATEs and pairwise average treatment effects among those receiving a specific treatment (average treatment effect on the treated, \cite{lopez-15}, \cite{mccaffrey-13}). We show that this method is generally valid and produces estimates that are relatively accurate and precise. 

\section{Notation for multiple treatments}

For $Z$ possible treatment groups, let $W_{i}$ denote the treatment group identification for unit $i$, where $W_{i}\in\mathcal{W}=\{1,\dots,Z\}$ and $i=1,\dots,n<\infty$. Let $n_{w}$ be the sample size of treatment group $w$ such that $\sum_{w=1}^{Z}n_{w}=n$. We define $T_{iw}$, $w\in\{1,\dots,Z\}$, to be an indicator variable for each unit $i$, $i=1,\dots,n$, that is equal to 1 if $W_{i}=w$ and to 0 otherwise. Thus, unit $i$ has a set of indicator variables, $\{T_{i1},\ldots,T_{iZ}\}$, where only $T_{iW_i}=1$ and the rest of the indicators are equal to zero. Let $\mathcal{Y}_{i}=\{Y_{i}(1),\dots,Y_{i}(Z)\}$ be the set of potential outcomes for unit $i$, where $Y_{i}(w)$ is the potential outcome for unit $i$ if it was exposed to treatment $w$. In practice, only the potential outcome corresponding to the intervention that affected unit $i$ is observed. The other potential outcomes cannot be observed because they correspond to treatment assignments that did not occur \cite{rubin-74, rubin-78}. Assuming the Stable Unit Treatment Value Assumption \cite{rubin-80}, the observed outcome for unit $i$ can be written as $Y^{obs}_{i}=T_{i1}Y_{i}(1)+\cdots+T_{iZ}Y_{i}(Z)$. Because we cannot directly observe the causal effect for unit $i$, we need to rely on multiple units of which some are exposed to each of the other $Z-1$ possible treatments. For drawing causal inference there are variables that are unaffected by $W_{i}$: covariates $X_{i}=(X_{i1},\dots,X_{iP})\in\mathbb{X}$. 

With $Z$ treatments, possible estimands of interest are the pairwise population average treatment effects between treatments $j$ and $k$, $\tau_{jk}=E\left(Y(j)-Y(k)\right)$, for $(j,k)\in\mathcal{W}^{2}$ and $j\neq k$ \cite{lopez-15}. A possible extension of $\tau_{jk}$ would be to contrast treatments among a subset of units in the population receiving baseline treatment $t\in\mathcal{W}$ and obtain the population average treatment effect on the treated \cite{mccaffrey-13}, $\tau_{jk}^{t}=E\left(Y(j)-Y(k)\mid W=t\right)$, for $(j,k)\in\mathcal{W}^{2}$ and $j\neq k$.

The estimands $\tau_{jk}$ and $\tau_{jk}^{t}$ can be approximated using the sample average treatment effects:
\[
	\hat{\tau}_{jk}=\frac{1}{n}\sum_{i=1}^{n}\left(Y_{i}(j)-Y_{i}(k)\right),\quad \hat{\tau}_{jk}^{t}=\frac{1}{n_{t}}\sum_{i=1}^{n}T_{it}\left(Y_{i}(j)-Y_{i}(k)\right).
\]
Because only one of $Y_{i}(w)$, $w\in\mathcal{W}$ is observed, matching procedures were proposed to impute the unobserved potential outcomes. For the remainder of the article, we will focus on estimating $\tau_{jk}^{t}$. In the Supplementary Material we present the corresponding results for $\tau_{jk}$. 

We assume the following convenient regularity condition: 

\begin{enumerate}
	\item[] \textit{Assumption 1}: $X_{i}$ is a random vector of continuous covariates distributed on $\mathbb{R}^{P}$ with compact and convex support $\mathbb{X}$, with density bounded away from zero on its support.
\end{enumerate}

Although Assumption 1 requires that all of the variables in $\mathbb{X}$ have a continuous distribution, discrete covariates with a finite number of support points can be accommodated by estimation of the treatment effects within subsamples defined by the values of these variables. 

A crucial piece of information that is needed for causal effect estimation is the assignment mechanism, or in other words, the probability for each unit to receive one of the $Z$ treatments, $P(W_{i}=w\mid X_{i},\mathcal{Y}_{i})$:

\begin{enumerate}
	\item[] \textit{Assumption 2}: Strong unconfoundedness and overlap \cite{lopez-15}
	\begin{enumerate}
		\item[] i. $P(W_{i}=t\mid X_{i},\mathcal{Y}_{i},\phi)=P(W_{i}=t\mid X_{i},\phi)\equiv r(t,X)$, where $\phi$ is a vector of parameters controlling the conditional distribution of $W_{i}=t$ and is notationally suppressed in $r(t,X)$. 
		\item[] ii. $r(t,X)<1-\eta$ for all $w\in\mathcal{W}$ and some $0<\eta<1$. 
	\end{enumerate}
\end{enumerate}

Under strong unconfoundedness, comparing individuals with similar $R(X)=(r(1,X),\dots,r(Z,X))$ results in well-defined causal effects \cite{imai-04}. Commonly, $R(X)$ is unknown and only an estimate of it is available, $\hat{R}(X) = (\hat{r}(1,X),\ldots,\hat{r}(Z,X))$.

Because treatment groups may be sampled separately and their sample sizes may not be proportional to their sizes in the population, we assume that sampling is random conditional on $W_{i}$. We also assume that each $n_{w}$, $w\neq t$, is at least the same order of magnitude as $n_{t}$. Formally:

\begin{enumerate}
	\item[] \textit{Assumption 3}: Conditional on $W_{i}=w$, the sample consists of independent draws from $Y,X\mid W=w$ for $w\in\mathcal{W}$. For some $r\geq1$, $n_{t}^{r}/n_{w}\to\rho$ with $0<\rho<\infty$.  
\end{enumerate} 

We also assume regularity conditions on the conditional moments of $Y(w)\mid X,W$:

\begin{enumerate}
	\item[] \textit{Assumption 4}: For covariates $x\in\mathbb{X}$ and treatment $w\in\mathcal{W}$, define $\mu_{w}(x)=E(Y(w)\mid X=x)$ and $\sigma_{w}^{2}(x)=Var(Y(w)\mid X=x)=Var(Y\mid X=x,W=w)$. Then, (i) $\mu_{w}(x)$ and $\sigma^{2}_{w}(x)$ are Lipschitz in $\mathbb{X}$ for all $w\in\mathcal{W}$, (ii) $E[(Y_{i}(w))^{4}\mid X_{i}=x]\leq C$ for some $C<\infty$, for almost all $x\in\mathbb{X}$, and (iii) $\sigma_{w}^{2}(x)$ is bounded away from zero. 
\end{enumerate} 

\section{The matching estimator}\label{sec:estimator}

\subsection{Point estimates}\label{subsec:standard}

For a vector $x \in X$, let $||x||_{A} = (x'Ax)^{1/2}$ for some positive definite matrix $A$. For example, when $A$ is the identity matrix this measure is the Euclidean distance. Our derivations will focus on matching with replacement, such that each unit can be used as a match more than once, and on the distance measure between units $i$ and $j$ of the form $||X_i -X_j||_A$. When matching on continuous covariates, matches are usually inexact, which generates bias in matching estimators. Matching with replacement increases the set of possible matches, which typically produces smaller biases \cite{abadie-06}. Let $\mathcal{M}_{i}^{w}$ denote the set of indices for the ``closest" $m$ units to unit $i$ that were exposed to treatment $w\neq W_{i}$, and $n_{w}\geq m$ for all $w$. Formally, $\sum_{j:W_j=w} \mathbb{I}\{||X_j - X_i|| \leq ||X_{l \in M_{i}^{W}} - X_i||\} = m$, where $\mathbb{I}$ is an indicator function, , equal to 1 if the expression in brackets is true and zero otherwise.

The matching estimator imputes the missing potential outcomes as the average of the $m$ observed outcomes of the units in $\mathcal{M}_{i}^{w}$, 
\[
	\hat{Y}_{i}(w)=
	\begin{cases}
	Y_{i}^{obs},& T_{iw}=1\\
	\frac{1}{m}\sum_{j\in\mathcal{M}_{i}^{w}}Y_{j}^{obs},& T_{iw}=0.
	\end{cases}
\]

Let $\psi_{iw}=\sum_{W_{j}=w}\mathbb{I}\{i\in\mathcal{M}_{j}^{W_{i}}\}$ be the number of times that unit $i$ serves as a match to other units in treatment group $w$, with $\psi_{iW_{i}}=0$. The point estimate for $\tau_{jk}^{t}$ is
\[
	\hat{\tau}_{jk}^{t}=\frac{1}{n_{t}}\sum_{W_{i}=t}\left(\hat{Y}_{i}(j)-\hat{Y}_{i}(k)\right)=\frac{1}{n_{t}}\sum_{i=1}^{n}(T_{ij}-T_{ik})\left(T_{it}+\frac{\psi_{it}}{m}\right)Y_{i}^{obs}, 
\]
We define $\tau^{t}=\{\tau_{jk}^{t}:j\in\mathcal{W}, k\in\mathcal{W}, j<k\}$ and its point estimate $\hat{\tau}^{t}_{M}=\{\hat{\tau}_{jk}^{t}:j\in\mathcal{W},k\in\mathcal{W},j<k\}$. The point estimate for $\tau_{jk}$ can be obtained by summing the $\hat{\tau}_{jk}^{t}$s of all reference groups $t$, weighted by their respective shares in the sample (see Supplementary Material).

The expectations of the potential outcomes are estimated by the sample averages, 
\[
	\tilde{Y}^{t}(w)\equiv\frac{1}{n_{t}}\sum_{W_{i}=t}\hat{Y}_{i}(w)=\frac{1}{n_{t}}\sum_{i=1}^{n}T_{iw}\left(T_{it}+\frac{\psi_{it}}{m}\right)Y_{i}^{obs}.
\]
Let $\mu_{w}^{t}=E(\mu_{w}(X)\mid W=t)$. We can decompose the bias $\tilde{Y}^{t}(w)-\mu_{w}^{t}$ into the following three parts \cite{abadie-06}:
\begin{align*}
	\tilde{Y}^{t}(w)-\mu_{w}^{t}&=\left(\overline{\mu_{w}}(X)-\mu_{w}^{t}\right)+B_{w}^{t}+E_{w}^{t},
\end{align*}
where $\overline{\mu_{w}}(X)\equiv \frac{1}{n_{t}}\sum_{i=1}^{n}T_{it}\mu_{w}(X_{i})$,
\begin{align*}
	B_{w}^{t}&\equiv \frac{1}{n_{t}}\sum_{i=1}^{n}\left[\frac{T_{it}(1-T_{iw})}{m}\sum_{j\in\mathcal{M}_{i}^{w}}\left(\mu_{w}(X_{i})-\mu_{w}(X_{j})\right)\right], \quad E_{w}^{t}\equiv\frac{1}{n_{t}}\sum_{i=1}^{n}T_{iw}\left(T_{it}+\frac{\psi_{it}}{m}\right)\epsilon_{iw}, 
\end{align*}
and $\epsilon_{iw}=Y_{i}(w)-\mu_{w}(X_{i})$. The terms $\left(\overline{\mu_{w}}(X)-\mu_{w}^{t}\right)$ and $E_{w}^{t}$ each have zero mean, and are asymptotically normal (see Section~\ref{subsec:normal}). 

If the number of continuous covariates $P\geq1$, then the conditional bias term $B_{w}^{t}$ is in general not $n^{1/2}$ consistent. Though if the parameter of interest is $\tau_{jk}^{t}$, the bias can be ignored if each $n_{w}$, $w\neq t$, is of sufficient order of magnitude \cite{abadie-06}.  
	
\subsection{Bias-corrected point estimates}\label{subsec:biascorrected}

To reduce the bias in point estimation that is a consequence of matching on a large number of covariates or when treatment $t$ is over-sampled relative to treatment $w\neq t$, we propose a regression imputation method. Let $\hat{\mu}_{w}(x)$ be a consistent estimator of $\mu_{w}(x)$. The bias-corrected matching estimator is
\[
	\hat{Y}_{i}^{\text{bc}}(w)=
	\begin{cases}
	Y_{i}(w),& T_{iw}=1\\
	\frac{1}{m}\sum_{j\in\mathcal{M}_{i}^{w}}\left(Y_{j}(w)+\hat{\mu}_{w}(X_{i})-\hat{\mu}_{w}(X_{j})\right),& T_{iw}=0.
	\end{cases}
\]
A possible estimate for $\hat{\mu}_{w}(X_{i})$ is $\hat{E}(Y_{i}^{obs}|X_{i},W_{i},\beta_{w})$ in each of the $Z$ treatment groups, estimated using multiple linear regression including all pretreatment covariates and possible second-order interactions. These estimators were shown to be robust to the misspecification of the regression function and are $n^{1/2}$ consistent, when using a nonparametric series estimator for $\mu_{w}(x)$ \cite{abadie-11}. Let $\hat{\tau}^{\text{bc},t}_{M}$ be $\hat{\tau}^{t}_{M}$ with $\hat{Y}_{i}(w)$ replaced by $\hat{Y}^{\text{bc}}_{i}(w)$, and let
\[
	 \hat{B}_{w}^{t}=\frac{1}{n_{t}}\sum_{i=1}^{n}\left[\frac{T_{it}(1-T_{iw})}{m}\sum_{j\in\mathcal{M}_{i}^{w}}\left(\hat{\mu}_{w}(X_{i})-\hat{\mu}_{w}(X_{j})\right)\right].
\]
The bias corrected matching estimator for $\tau_{jk}^{t}$ is $\hat{\tau}_{jk}^{\text{bc},t}=\hat{\tau}_{jk}^{t}-(\hat{B}_{k}^{t}-\hat{B}_{j}^{t})$. Based on Theorem 2' in Abadie and Imbens \cite{abadie-11}, $\hat{\tau}^{\text{bc},t}_{M}$ has the same asymptotic variance as $\hat{\tau}^{t}_{M}$. 

\subsection{Sampling variance}\label{subsec:variance}

In order to calculate the marginal sampling variance of $\hat{\tau}_{M}^{t}$, it is useful to decompose it into a linear combination of the vector $\tilde{Y}^{t}=(\tilde{Y}^{t}(1),\dots,\tilde{Y}^{t}(Z))$, such that 
\[
	\hat{\tau}_{M}^{t}=
	\begin{pmatrix}
	\frac{1}{n_{t}}\sum_{W_{i}=t}\left(\hat{Y}_{i}(1)-\hat{Y}_{i}(2)\right)\\
	\frac{1}{n_{t}}\sum_{W_{i}=t}\left(\hat{Y}_{i}(1)-\hat{Y}_{i}(3)\right)\\
	\vdots\\
	\frac{1}{n_{t}}\sum_{W_{i}=t}\left(\hat{Y}_{i}(Z-1)-\hat{Y}_{i}(Z)\right)
	\end{pmatrix}=
	\begin{pmatrix}1&-1&0&\cdots&0&0\\1&0&-1&\cdots&0&0\\\vdots&\vdots&\vdots&\ddots&\vdots&\vdots\\0&0&0&\cdots&1&-1
	\end{pmatrix}
	\begin{pmatrix}
	\tilde{Y}^{t}(1)\\\tilde{Y}^{t}(2)\\\vdots\\\tilde{Y}^{t}(Z)\end{pmatrix}
	=A\tilde{Y}^{t}.
\]

\begin{lemma}
If Assumption 3 holds, then $\tilde{Y}^{t}(w)$ and $\tilde{Y}^{t}(w')$ are independent for all $w\neq w'$. 
\end{lemma}

(Proof: see Supplementary Material.)

Based on Lemma 1, $Cov(\tilde{Y}^{t}(w),\tilde{Y}^{t}(w'))=0$ for $w\neq w'$, thus for $j\neq k\neq l$:
\begin{align*}
	Cov(\tilde{Y}^{t})&=diag\left(Var(\tilde{Y}^{t}(1)),\dots,Var(\tilde{Y}^{t}(Z))\right),\\
	Cov(\hat{\tau}_{jk}^{t},\hat{\tau}_{jl}^{t})&=Var(\tilde{Y}^{t}(j))\\
	Var(\hat{\tau}_{M}^{t})&= A\{Cov(\tilde{Y}^{t})\}A^{\text{T}}
\end{align*}

Using the law of total variance, the marginal variance of $\tilde{Y}^{t}(w)$ can be expressed as
\begin{equation}
	Var(\tilde{Y}^{t}(w))=\frac{E\{n_{t}Var(\tilde{Y}^{t}(w)\mid X,W)\}+Var(\mu_{w}(X))}{n_{t}}, 
	\label{var.Yt}
\end{equation}
where
\begin{equation}
	Var(\tilde{Y}^{t}(w)\mid X,W)=\frac{1}{n_{t}^{2}}\sum_{i=1}^{n}T_{iw}\left(T_{it}+\frac{\psi_{it}}{m}\right)^{2}\sigma^{2}_{W_{i}}(X_{i}).
	\label{cond.var.Yt}
\end{equation}

Using Equations~\ref{var.Yt}--\ref{cond.var.Yt}, and the derivations given in Abadie and Imbens \cite{abadie-06}, pages 250-251, the estimated marginal variance of $\hat{\tau}_{jk}^{t}$ is 
\[
	\widehat{Var}(\hat{\tau}_{jk}^{t})=\frac{1}{n_{t}^{2}}\sum_{W_{i}=t}\left(\hat{Y}_{i}(j)-\hat{Y}_{i}(k)-\hat{\tau}_{jk}^{t}\right)^{2}+\frac{1}{n_{t}^{2}}\sum_{i=1}^{n}(T_{ij}+T_{ik})\left(\frac{\psi_{it}(\psi_{it}-1)}{m^{2}}\right)\hat{\sigma}^{2}_{W_{i}}(X_{i}),
\]
where $\hat{\sigma}_{W_{i}}^{2}(X_{i})$ is an estimate of the conditional outcome variance, $\sigma_{W_{i}}^{2}(X_{i})$. 

Abadie and Imbens \cite{abadie-06} showed that, with binary treatment, $\widehat{Var}(\hat{\tau}_{12}^{1})$ is consistent. The following theorem shows that the estimated marginal covariance between $\hat{\tau}_{jk}^{t}$ and $\hat{\tau}_{jl}^{t}$ is consistent for $Var(\tilde{Y}^{t}(j))$. 

\begin{theorem}
Let $\hat{\sigma}_{W_{i}}^{2}(X_{i})$ be an estimate of the conditional outcome variance, $\sigma_{W_{i}}^{2}(X_{i})$. Then
\[
	\widehat{Cov}(\hat{\tau}_{jk}^{t},\hat{\tau}_{jl}^{t})=\widehat{Var}(\tilde{Y}^{t}(j))=\frac{1}{n_{t}^{2}}\sum_{W_{i}=t}\left(\hat{Y}_{i}(j)-\tilde{Y}^{t}(j)\right)^{2}+\frac{1}{n_{t}^{2}}\sum_{i=1}^{n}T_{ij}\left(\frac{\psi_{it}(\psi_{it}-1)}{m^{2}}\right)\hat{\sigma}^{2}_{W_{i}}(X_{i}).
\]
If Assumptions 1--4 hold, then $\widehat{Var}(\tilde{Y}^{t}(j))$ is consistent for $Var(\tilde{Y}^{t}(j))$. 
\end{theorem}

(Proof: see Supplementary Material.)

Because $\widehat{Cov}(\hat{\tau}_{jk}^{t},\hat{\tau}_{jl}^{t})$ is consistent for $Var(\tilde{Y}^{t}(j))$, it follows that we can consistently estimate $Cov(\tilde{Y}^{t})$. Therefore, we can consistently estimate $Var(\hat{\tau}_{M}^{t})$ using a matrix consisting of marginal variances of the $\tilde{Y}^{t}(j)$s. 

\subsection{Estimating $\sigma_{w}^{2}(X)$}

For binary treatment, Abadie and Imbens \cite{abadie-06} developed a procedure that estimates $\sigma_{w}^{2}(X)$ by matching unit $i$ in treatment group $w$ to the closest unit(s) in treatment group $w$, in terms of the propensity score. We extend this procedure by finding units with similar $R(X)$. 

Let $\mathcal{L}_{i}^{w}$ be the set of matches for unit $i$ from its own treatment group $w$, excluding unit $i$ itself. Let $\hat{d}(w,X) = \log\left(\frac{\hat{r}(w,X)}{\hat{r}(Z,X)}\right)$ and $\hat{D}(X) = (\hat{d}(1,X),\ldots,\hat{d}(Z,X))$. When $R(X)$ is estimated using multinomial logistic regression, $\hat{d}(w,X)$ is a linear combination of $X$, $\hat{\gamma}'X$. We define the distance between two units to be the Euclidean distance of $\hat{D}(X)$: $\sum_{w \in{1,...,Z-1}}\left\{\text{logit}(\hat{r}(w,X_{i'}))-\text{logit}(\hat{r}(w,X_{i}))\right\}^{2}$, where $i'=1,\dots,n_{w}$ and $i\neq i'$. 

When $|\mathcal{L}_{i}^{w}|=J$, $\sigma_{w}^{2}(X_{i})$ can be estimated by
\[
	\hat{\sigma}_{w}^{2}(X_{i})=\frac{J}{J+1}\left(Y_{i}^{obs}-\frac{1}{J}\sum_{j=1}^{J}Y_{\ell_{j}^{w}(i)}^{obs}\right)^{2},
\]
where $\ell_{j}^{w}(i)$ is the index of the $j$th closest match to unit $i$ from treatment group $w$.

If the covariate values for unit $i$ are equal to the covariate values of all units in $\mathcal{L}_{i}^{w}$, then $\hat{\sigma}_{w}^{2}(X_{i})$ is an unbiased estimate of $\sigma_{w}^{2}(X_{i})$. However, it is not a consistent estimate for $\sigma_{w}^{2}(X_{i})$, though appropriately weighted averages of the $\hat{\sigma}^{2}_{w}(X_{i})$ over the sample are consistent for $Var(\tilde{Y}(w)\mid X,W)$ and $n_{t}Var(\tilde{Y}^{t}(w)\mid W,X)$ \cite{abadie-06}. 

Generally, the covariate values differ between units. Imbens and Rubin \cite{imbens-15} proposed a bias corrected version of $\hat{\sigma}^{2}_{w}(X_{i})$ for binary treatment settings. The procedure relies on the regression model $E(Y_{i}^{obs}\mid X_{i},W_{i}=w)=X_{i}\beta_{w}$ for each treatment group, and $\hat{\sigma}^{2}_{w}(X_{i})$ is estimated by the variance of the residuals obtained from these regression models for unit $i$ and its $J$ closest matches. A similar method can be implemented for multiple treatments, by estimating separate regression models in each of the $Z$ treatment groups. 

\subsection{Asymptotic normality}
\label{subsec:normal}

We state the formal result for asymptotic normality of the $\tilde{Y}^{t}(w)$s after subtracting the conditional bias term. 
\begin{theorem}~
\begin{enumerate}
	\item[] i. Suppose Assumptions 1--4 hold. Then 
	\[
		[Var(\tilde{Y}^{t}(w))]^{-1/2}\left(\tilde{Y}^{t}(w)-B_{w}^{t}-\mu_{w}^{t}\right)\to N(0,1), 
	\]
	and $[Var(\hat{\tau}_{M}^{t})]^{-1/2}(\hat{\tau}_{M}^{t}-B_{M}^{t}-\tau^{t})\to N_{p}(0,I_{p})$. \mbox{}\\
	\item[] ii. Suppose Assumptions 1--4 hold. Then under several additional regularity conditions (see Supplementary Material), 
	\[
		[Var(\tilde{Y}^{t}(w))]^{-1/2}\left(\tilde{Y}^{\text{bc},t}(w)-\mu_{w}^{t}\right)\to N(0,1), 
	\]
	and $[Var(\hat{\tau}_{M}^{t})]^{-1/2}(\hat{\tau}_{M}^{\text{bc},t}-\tau^{t})\to N_{p}(0,I_{p})$.
\end{enumerate}
\end{theorem}

(Proof: see Supplementary Material.)
 
Based on Theorem 3.3, 
\[
	z^{2}=(\hat{\tau}_{M}^{t}-B_{M}^{t}-\tau^{t})^{\text{T}}[Var(\hat{\tau}_{M}^{t})]^{-1}(\hat{\tau}_{M}^{t}-B_{M}^{t}-\tau^{t})\sim\chi^{2}_{p},
\]
and a $100(1-\alpha)\%$ confidence region is the region such that $P\left(z^{2}\leq\chi^{2}_{p}\right)=1-\alpha$ \cite{siddique-18}.

\subsection{Weighting Estimators}

Inverse probability weighting (IPW) is another common approach for estimating causal effects with multiple treatments \cite{mccaffrey-13, siddique-18, feng-11}. Under weak unconfondedness \cite{imbens-00}, $\tau^{t}_{jk}$ can be estimated as 
\begin{align}
	\hat{\tau}_{jk}^{t,IPW}&=E(\hat{Y}_{i}(j))-E(\hat{Y}_{i}(k))\nonumber\\
	&= \left[\left(\sum_{i=1}^{n}\frac{\mathbb{I}(W_{i}=j)Y_{i}^{obs}r(t,X_{i})}{r(j,X_{i})}\right)\times\left(\sum_{i=1}^{n}\frac{\mathbb{I}(W_{i}=j)r(t,X_{i})}{r(j,X_{i})}\right)^{-1}\right]\nonumber\\
	& \ \ \ \ -  \left[\left(\sum_{i=1}^{n}\frac{\mathbb{I}(W_{i}=k)Y_{i}^{obs}r(t,X_{i})}{r(k,X_{i})}\right)\times\left(\sum_{i=1}^{n}\frac{\mathbb{I}(W_{i}=k)r(t,X_{i})}{r(k,X_{i})}\right)^{-1}\right]\label{ipw}.
\end{align}

Because $R(X)$ is commonly unavailable, it is replaced by $\hat{R}(X)$ in Equation~\ref{ipw}. Thus, in order to calculate the sampling variance of this estimate, the sandwich estimator, which takes into account the uncertainty in the estimated propensity score, is commonly implemented \cite{mccaffrey-13, siddique-18}. 

The consistency of IPW estimates relies on the correct specification of the GPS model. A ``doubly robust" estimator attempts to overcome this limitation by  combining weighting with regression adjustments. Formally, $E(\hat{Y}_{i}(j))$ in Equation~\ref{ipw} is replaced by $n^{-1}\sum_{i=1}^{n}\left(\frac{T_{ij}Y_{i}^{obs}}{r(j,X_i)} - \frac{(T_{ij}-r(j,X_i))}{r(j,X_i)}\hat{\mu}_{w}(X_i)\right)$.  

The ``doubly robust" estimator yields consistent estimates of the treatment effect if either the model for the outcome or the propensity score model is correct, but not necessarily both \cite{kang-07}.

\section{Simulations}

\subsection{Evaluating coverage of matching estimator by simulation}

We examined the operating characteristics of the different procedures in finite samples, using simulations. The derivations in Section \ref{sec:estimator} are limited to a specific matching algorithm that relies on the $||X_{i}-X_{j}||_{A}$ distance measure between units $i$ and $j$. Thus, we examined the performance of the matching estimators using either the vector matching algorithm \cite{lopez-15} or a matching algorithm that is based on the Mahalanobis distance of the logit GPS vector \cite{scotina-18}. Because both matching algorithms yielded similar results, we provide the results for vector matching in the manuscript and the results for matching on the Mahalanobis distance of the logit GPS vector are described in the Supplementary Material (Table 1). The simulations were conducted using the Matching package \cite{sekhon-11} in R Studio \cite{RSoftware}. 

The performance of the different matching estimators were compared using a complete factorial design. The simulation configurations comprise two types of factors. The first set of factors describes the covariate distributions and sample sizes, which are either known to the investigator, or can be easily estimated without examining outcome data. The second set of factors involves the response surfaces which are unknown to the investigator and cannot be estimated at the design stage. The covariates' values of the $n_{1}$, $n_{2}=\gamma n_{1}$, and $n_{3}=\gamma^{2}n_{1}$ units receiving treatments 1, 2, and 3, respectively, are generated from multivariate normal or multivariate $t_{7}$ distributions:
\begin{align*}
	X_{i}\mid\{W_{i}=1\}&\sim f(\mu_{1},\Sigma_{1}),\ i=1,\dots,n_{1},\\
	X_{i}\mid\{W_{i}=2\}&\sim f(\mu_{2},\Sigma_{2}),\ i=n_{1}+1,\dots,n_{1}+\gamma n_{1},\\
	X_{i}\mid\{W_{i}=3\}&\sim f(\mu_{3},\Sigma_{3}),\ i=n_{1}+\gamma n_{1}+1,\dots, n_{1}+\gamma n_{1}+\gamma^{2}n_{1},
\end{align*}
where $f\in\{N, t_{7}\}$, and 
\begin{align*}
	\mu_{1}&=((b,0,0),\dots,(b,0,0))^{\text{T}},\\
	\mu_{2}&=((0,b,0),\dots,(0,b,0))^{\text{T}},\\
	\mu_{3}&=((0,0,b),\dots,(0,0,b))^{\text{T}}.
\end{align*}
The covariance matrices $\Sigma_{1}$, $\Sigma_{2}$, and $\Sigma_{3}$ have respective diagonal entries of $1$, $\sigma_{2}^{2}$, and $\sigma_{3}^{2}$, and $\lambda$ elsewhere. Additionally, we examined configurations in which $X_{i3}$ and $X_{i6}$ were converted to binary variables, such that $X_{ip}=1$ if $X_{ip}>0$ and 0 otherwise, $p\in\{3,6\}$. 

For each configuration of the design factors, we generate the potential outcomes as $Y_{i}(w)=\sum_{p=1}^{P}g(X_{ip})^{\text{T}}\beta_{wp}+\epsilon_{w,i}$, where $\beta_{wp}\overset{iid}{\sim}\text{Uniform}(-\theta, \theta)$, and $\epsilon_{w,i}\sim N(0,1)$ for $i=1,\dots,n$. These response surfaces imply that the treatment effects are correlated with $X$ and are non-additive. 

\begin{table}[!htb]
\centering
\caption{Simulation factors}
\begin{tabular}{l l}
\toprule
Factor & Levels of factor\\
\midrule
$n_{1}$ & $\{600, 1200\}$\\
$\gamma=\frac{n_{2}}{n_{1}}=\frac{n_{3}}{n_{2}}$ & $\{1,2\}$\\
$b$ & $B=\frac{b}{\sqrt{\frac{1+\sigma_{2}^{2}+\sigma_{3}^{2}}{3}}}\ \text{takes levels}\ \{0, 0.25, 0.50, 0.75, 1.00\}$\\
$\lambda$ & $\{0, 0.25\}$\\
$\sigma_{2}^{2}$ & $\{0.5, 1, 2\}$\\
$\sigma_{3}^{2}$ & $\{0.5, 1, 2\}$\\
$\theta$ & $\{2, 5, 10\}$\\
$P$ & $\{3, 6\}$\\
$f$ & $\{N, t_{7}\}$\\
$g$ & $\{X, \exp(X/10)\}$\\
\bottomrule
\end{tabular}
\label{sim.factors}
\end{table}

The distributions of $X$ and $Y(w)$ are varied by ten factors, resulting in a $2^{6}\times 3^{3}\times 5$ factorial design (Table~\ref{sim.factors}). For each configuration, 500 replications were produced. We evaluate the performance of the basic (Section \ref{subsec:standard}) and bias-corrected (Section \ref{subsec:biascorrected}) point estimates, and their combination with the newly proposed standard error (Section \ref{subsec:variance}) and with randomization based standard error \cite{austin-09}. We denote these as B-N, BC-N, B-R, and BC-R, where B stands for the basic point estimate, BC stands for the bias-corrected point estimate, and N and R stand for the newly proposed and the randomization based standard errors, respectively. For all matching procedures, we used $m=1$, and for estimating the conditional outcome variance we assumed that $J=1$.

We also evaluate the performance of inverse probability weighting (IPW) and doubly robust (DR) estimators \cite{mccaffrey-13}, which estimate each $\tau_{jk}^{1}$ separately. For each replication, we calculated the estimated treatment effects, the estimated sampling covariance matrices, the corresponding 95\% confidence regions and 95\% Bonferroni-adjusted pairwise confidence intervals \cite{dunn-61}, and determined whether these regions and intervals covered the treatment effects. Results for configurations with only continuous covariates show similar trends and are reported in the Supplementary Material (Table 2). 

\subsection{Results for 95\% region and interval coverages}

Using simulations, we demonstrate that the basic and bias-corrected point estimates, in combination with the newly proposed sampling variance estimate, result in statistically valid methods, while a combination of these point estimates with randomization based sampling variance are generally invalid. Table~\ref{coverage.table} displays the median and interquartile range of the coverages, with confidence regions as well as using Bonferroni-corrected pairwise intervals. The basic and bias-corrected matching estimators with the proposed standard error estimates (B-N and BC-N) have median confidence region coverages that are at or higher than nominal. The basic and bias-corrected matching estimators with randomization based standard error estimates (B-R and BC-R) have median coverage region that is less than 0.75 for a 95\% confidence region.

\begin{table}[!h]
\centering
\caption{Median, 25\% percentile, and 75\% percentile of the 95\% region coverage for $\tau^{1}$, and the 95\% interval coverage for $\tau_{jk}^{1}$ averaged over three estimands (B: basic point estimate; BC: bias-corrected point estimate; N: newly proposed standard errors; R: randomization based standard errors; IPW: inverse probability weighting; DR: doubly robust estimation)}
\begin{tabular}{l l l l l l l l l l l l l l}
\toprule
& & \multicolumn{5}{c}{Confidence Region} & & & \multicolumn{5}{c}{Confidence Interval}\\
& & Median & & 25\% & & 75\% & & & Median & & 25\% & & 75\%\\
\midrule
B-N & & 0.95 & & 0.78 & & 0.98 & & & 0.96 & & 0.86 & & 0.98\\
BC-N & & 0.97 & & 0.87 & & 0.99 & & & 0.98 & & 0.92 & & 0.99\\
B-R & & 0.70 & & 0.33 & & 0.88 & & & 0.81 & & 0.57 & & 0.92\\
BC-R & & 0.75 & & 0.34 & & 0.91 & & & 0.86 & & 0.62 & & 0.95\\
IPW & & -- & & -- & & -- & & & 0.85 & & 0.57 & & 0.95\\
DR & & -- & & -- & & -- & & & 0.85 & & 0.57 & & 0.94\\
\bottomrule
\end{tabular}
\label{coverage.table}
\end{table}

B-R, BC-R, IPW, and DR, have median interval coverage that is lower than nominal. Among these methods, the 25th percentile of coverage is the highest for BC-N. B-N and BC-N have median interval coverages that are at or greater than nominal. 

To identify the factors with the largest influence on the coverage for each of the methods, we order them by their mean squared error for coverage rate (as in \cite{rubin-79} and \cite{cangul-09}). The number of covariates ($P$), the ratio of units receiving $W=2$ to those receiving $W=1$ ($\gamma$), the initial covariate bias ($b$), and their interactions explain 61\%, 58\%, 83\%, and 70\% of the variability in coverage for B-N, BC-N, B-R, and BC-R, respectively. For IPW and DR, $b$ and $\gamma$ explained 32\% and 29\% of the variability in coverage, respectively. The number of covariates $P$ was not as influential when using IPW or DR.

\begin{table}[!h]
\centering
\caption{Median region coverage for equal and unequal sample sizes}
\begin{tabular}{l c c c c c c c c c c c c}
\toprule
& & \multicolumn{2}{c}{B-N} && \multicolumn{2}{c}{BC-N} && \multicolumn{2}{c}{B-R} && \multicolumn{2}{c}{BC-R}\\
& $b$ & $P=3$ & $P=6$ && $P=3$ & $P=6$ && $P=3$ & $P=6$ && $P=3$ & $P=6$\\
\midrule
\multirow{5}{*}{$\gamma=1$}
& 0.00 & 0.97 & 0.98 && 0.98 & 0.99 && 0.89 & 0.89 && 0.90 & 0.93\\
& 0.25 & 0.96 & 0.94 && 0.98 & 0.97 && 0.77 & 0.73 && 0.79 & 0.76\\
& 0.50 & 0.93 & 0.86 && 0.96 & 0.94 && 0.56 & 0.43 && 0.51 & 0.48\\
& 0.75 & 0.87 & 0.77 && 0.93 & 0.88 && 0.32 & 0.23 && 0.23 & 0.24\\
& 1.00 & 0.77 & 0.65 && 0.87 & 0.82 && 0.14 & 0.10 && 0.08 & 0.09\\
\midrule
\multirow{5}{*}{$\gamma=2$} 
& 0.00 & 0.97 & 0.98 && 0.97 & 0.99 && 0.95 & 0.94 && 0.96 & 0.96\\
& 0.25 & 0.93 & 0.98 && 0.96 & 0.99 && 0.87 & 0.92 && 0.91 & 0.95\\
& 0.50 & 0.83 & 0.98 && 0.90 & 0.99 && 0.69 & 0.84 && 0.78 & 0.90\\
& 0.75 & 0.72 & 0.97 && 0.78 & 0.99 && 0.45 & 0.67 && 0.51 & 0.77\\
& 1.00 & 0.55 & 0.96 && 0.65 & 0.99 && 0.22 & 0.50 && 0.25 & 0.57\\
\bottomrule
\label{coverage.g}
\end{tabular}
\end{table}

We further detail the effects of the principal determinants of coverage by averaging over the other factors. Table~\ref{coverage.g} shows the median region coverage for the newly proposed methods based on $b$, $P$, and $\gamma$. In settings with $\gamma=1$ and $P=3$, B-N and BC-N have median coverages that are at or above nominal across all levels of $b$. In settings with $\gamma=1$ and $P=6$, B-N and BC-N have median coverages that are at or above nominal for $b\leq0.5$, and lower than nominal when $b>0.5$. The latter configurations are hard cases for matching estimators \cite{rubin-79}. Both point estimates with randomization based sampling variance have close to nominal coverage only when the covariate distributions are similar across treatment groups ($b=0.00$), which is practically a randomized experiment. Median coverages are higher than nominal for B-N and BC-N when $\gamma=2$. Median coverages for IPW or DR were lower than nominal for $b\geq0.25$ and $t$-distributed continuous covariates (data in Table 3 of Supplementary Material). 

\subsection{Results for biases and interval widths}

The bias-corrected matching estimator has the lowest bias, and weighting methods have the largest biases for $b > 0$. The first part of Table \ref{bias.width} depicts the median absolute biases for the basic, bias-corrected, IPW, and DR point estimates across the levels of $b$. Because there are three point estimates, we calculated the overall bias by averaging the absolute bias of each of the three point estimates for each configuration, and then estimated the median across configurations and levels of $b$. When $b$ increases, the median overall absolute bias increases for all of the methods, with the bias-corrected point estimate having the smallest bias. For IPW and DR, the median and IQR absolute biases increase faster with $b$, and for $b\geq0.50$, they are over double the size of the biases obtained for the basic matching estimator.

\begin{table}[!h]
\centering
\caption{Median absolute bias, interval width, and ratio of the newly proposed, randomization based, IPW, and DR standard errors to the empirical standard errors averaged over the three estimates, across 8,640 configurations (parentheses are the interquartile range)}
\begin{tabular}{l l c c c c c}
\toprule
& & \multicolumn{5}{c}{$b$}\\
& Method & 0.00 & 0.25 & 0.50 & 0.75 & 1.00\\
\midrule
Bias & B & 0.03 (0.04) & 0.07 (0.16) & 0.12 (0.32) & 0.19 (0.52) & 0.27 (0.78)\\
& BC & 0.03 (0.04) & 0.06 (0.16) & 0.11 (0.30) & 0.17 (0.48) & 0.25 (0.70)\\
& IPW & 0.02 (0.03) & 0.10 (0.26) & 0.19 (0.51) & 0.29 (0.79) & 0.38 (1.04)\\
& DR & 0.02 (0.03) & 0.09 (0.25) & 0.18 (0.51) & 0.28 (0.78) & 0.39 (1.08)\\
\midrule
Interval & N & 0.63 (1.37) & 0.64 (1.42) & 0.75 (1.66) & 0.92 (1.97) & 1.20 (2.46)\\
Width & R & 0.49 (1.13) & 0.47 (1.08) & 0.47 (1.10) & 0.49 (1.14) & 0.52 (1.22)\\
& IPW & 0.38 (0.84) & 0.41 (0.91) & 0.53 (1.19) & 0.79 (1.83) & 1.18 (2.84)\\
& DR & 0.37 (0.77) & 0.40 (0.83) & 0.51 (1.05) & 0.74 (1.60) & 1.07 (2.40)\\
\midrule
SE & N & 1.02 & 1.04 & 1.05 & 1.05 & 1.06\\
Ratio & R & 0.81 & 0.76 & 0.67 & 0.55 & 0.45\\
& IPW & 1.13 & 1.14 & 1.16 & 1.07 & 0.96\\
& DR & 1.10 & 1.12 & 1.14 & 1.08 & 0.99\\
\bottomrule
\end{tabular}
\label{bias.width}
\end{table}

The second part of Table \ref{bias.width} depicts the median interval width using the newly proposed and the randomization based sampling variances, IPW and DR, across levels of $b$. We used Bonferroni correction to obtain the three 95\% pairwise confidence intervals for each configuration, and we examined the median of the average of the three interval widths for all configurations across the levels of $b$. The median interval widths are overall larger for the newly proposed standard errors than for the randomization based standard errors. However, the randomization based standard errors generally result in invalid statistical procedures. For IPW and DR, the median and IQR interval widths are smaller for $b\leq0.50$, and increase sharply for larger $b$.

The newly proposed sampling variance estimates approximate the empirical sampling variance well. We calculated the average of the ratios of the randomization based and newly proposed standard errors of $\hat{\tau}_{12}^{1}$, $\hat{\tau}_{13}^{1}$, and $\hat{\tau}_{23}^{1}$, to their empirical standard errors for each configuration, and did the same for the IPW and DR standard errors. The medians of the ratios for the newly proposed, IPW, and DR standard errors are close to 1 for all levels of $b$. These findings imply that the these standard error estimates provide good approximations to the empirical ones, and the poorer coverage rates for IPW and DR stem from large biases as a result of increasing $b$. The medians of these ratios for the randomization based standard errors are smaller than 1, implying that they underestimate the standard errors. 

\section{Real data application}

\subsection{Emergency department data}

We compare the effects of different pain medication regimens on long-term pain among patients that were discharged from the emergency department (ED) after motor vehicle collision (MVC). The data were collected as part of two large, multicenter, prospective cohort studies of adult patients who presented to an ED within 24 hours of a MVC and were discharged to home after evaluation \cite{beaudoin-18}. Here, we examined three medication regimens: opioid analgesic, non-steroidal anti-inflammatory drugs (NSAIDs), and a combination of both. The primary outcome was the self-reported overall pain severity six weeks after the MVC, which was assessed using a 0 to 10 numeric rating scale. GPS models were fit using multinomial logistic regression that included demographics characteristics, accident characteristics, initial pain, clinical characteristics and comorbidities. The pre-matched cohort included 257 patients that were prescribed opioids, 951 patients that were prescribed NSAIDs, and 110 patients that were prescribed both. Because some of the baseline covariates and the outcome suffered from missing values we created 20 multiple complete datasets using the fully conditional specification approach in each arm separately \cite{vanbuuren-07}. Each complete dataset was analyzed separately using the proposed approach and the results were combined using the common combination rules \cite{rubin-04}. The multiple imputation procedure was performed using the mice package in R \cite{su-11}. This analysis practically assumes that missing outcomes are not differentially missing across treatment arms and that they are not influenced by the unobserved potential outcomes. These are strong assumptions that should generally be examined using sensitivity analyses. 

\begin{figure}[!h]
\centering
\includegraphics[scale=0.5]{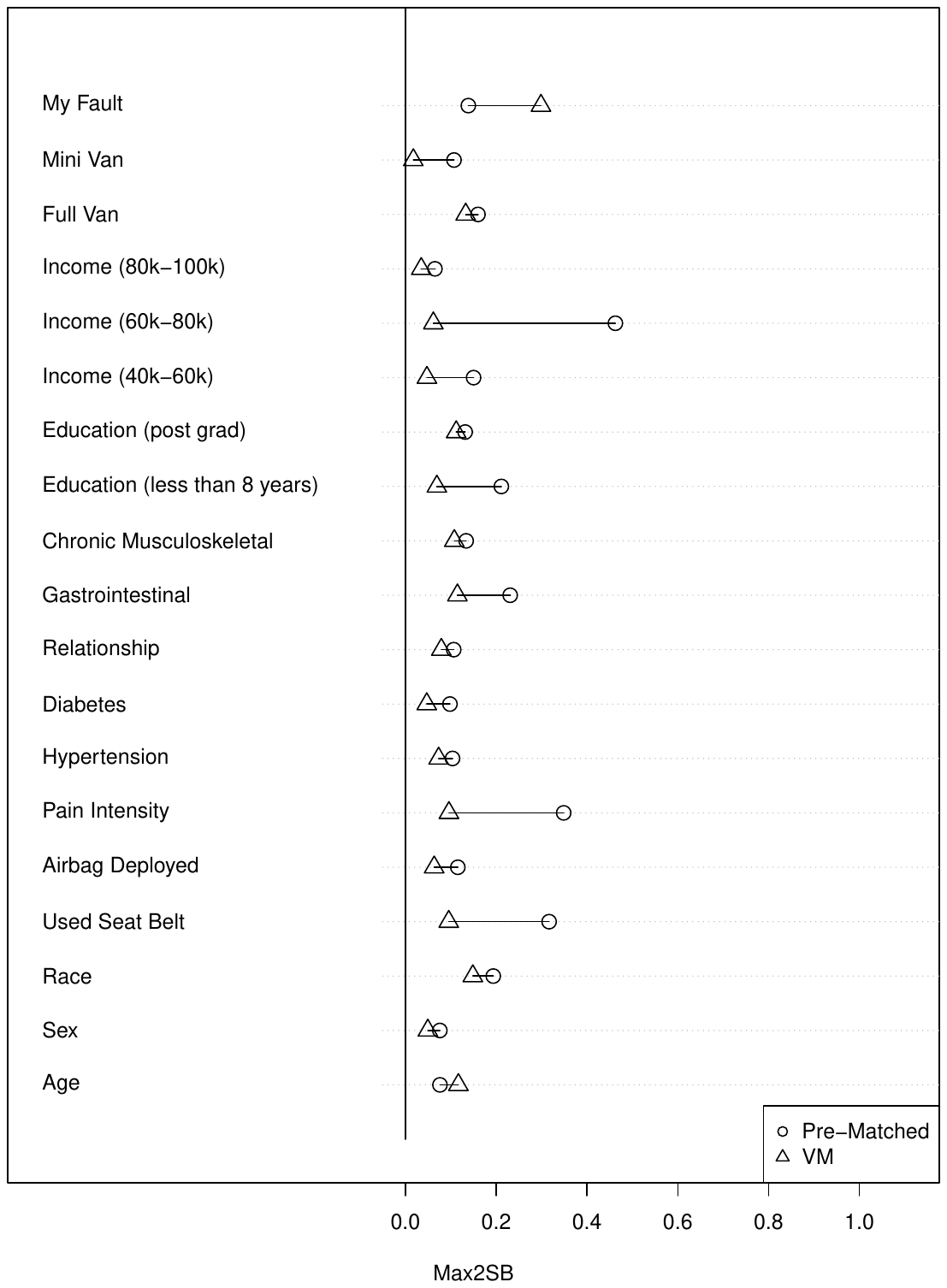}
\caption{$Max2SB_{p}$ for the pre-matched cohort and after vector matching}
\label{matching}
\end{figure}

Matching quality was assessed using diagnostics presented in McCaffrey et al. \cite{mccaffrey-13}, Lopez and Gutman \cite{lopez-15}, and Scotina and Gutman \cite{scotina-18}, for all of the imputed datasets. Vector matching retained 96\% of reference treatment patients in the matched cohort, and yielded a maximum absolute covariate pairwise bias of 0.29, as compared to 0.46 in the pre-matched cohort. Figure~\ref{matching} reports the maximum absolute standardized pairwise bias, $Max2SB_{p}$, for each of 18 of the baseline covariates in the original, unmatched sample, and for the matched sample for one of the imputed datasets. Vector matching yielded improved balance for most of the covariates.  Results are similar for each imputed dataset (data not shown).

\begin{figure}[!h]
\centering
\includegraphics[scale=0.5]{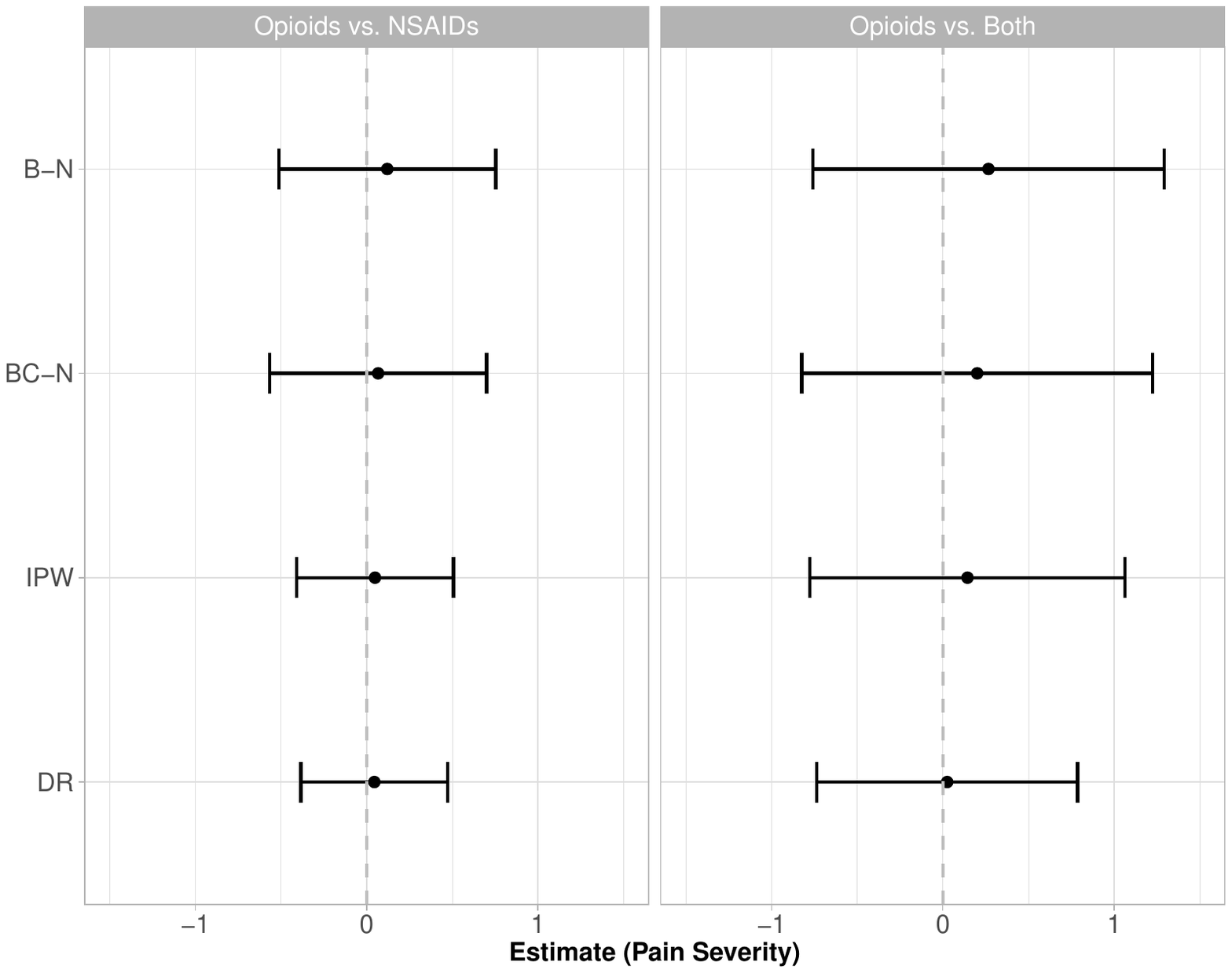}
\caption{Estimated average treatment effects and 95\% confidence intervals comparing patients receiving opioid treatment after MVC to patients receiving NSAID treatment, or concurrent opioid and NSAID treatment}
\label{ed.data}
\end{figure}

We performed simultaneous comparison across the three estimates by combining the $\chi^{2}$ statistics across the 20 multiply imputed datasets \cite{enders-10}. For B-N we obtained $\chi^{2}=0.18$ and $p$-value $=0.91$, and for BC-N we obtained $\chi^{2}=0.11$ and $p$-value $=0.95$. Thus, at the 5\% level, we are unable to reject the global null hypothesis of similar pain levels six weeks after MVC between the three medication regimens. Figure~\ref{ed.data} displays the estimated pairwise average treatment effects between patients receiving opioids versus NSAIDs, and opioids versus concurrent treatment with opioids and NSAIDs, among patients receiving opioids, using the B-N, BC-N, IPW, and DR. The estimated difference in reported pain severity 6 weeks after the MVC between opioid and NSAID users using BC-N was $0.07$ with 95\% confidence interval $(-0.56,0.70)$, similar results were observed for the other methods, with DR having the shortest interval . The estimated difference in reported pain severity 6 weeks after the MVC between opioid and concurrent opioid and NSAID users using BC-N was $0.20$ with 95\% confidence interval $(-0.83, 1.23)$, with similar results observed for the other methods.

\section{Discussion}

This paper proposes point and interval matching estimators for the average treatment effects and the average treatment effects on the treated with multiple nominal treatments. This method is an extension of the point and interval estimates developed by Abadie and Imbens \cite{abadie-06, abadie-11} for binary treatment. Our derivations can also be used to obtain point and interval estimates for other linear contrasts of the expectations of the potential outcomes. 

Using simulations, we demonstrate that the basic and bias-corrected point estimates, in combination with the newly proposed sampling variance estimate, result in statistically valid methods, while a combination of these point estimates with randomization based sampling variance are generally invalid. The IPW and DR point estimates result in statistically valid methods for small $b$ and normally distributed covariates, but are generally invalid for large $b$ and non-normally distributed covariates. In addition, the newly proposed sampling variance estimates approximate the empirical sampling variance well. 

The simulation study used $m=1$ and $J=1$ for all configurations because improvements in precision from using larger $m$ and $J$ are generally minimal in large samples, and can potentially increase covariates' bias \cite{imbens-15}. An area for future work would be to identify $m$ and $J$ with optimal operating characteristics. Another possible extension is to develop the point and interval estimates for binary and count outcomes. Lastly, an important area of future research is to evaluate the performance of matching estimators when the dimension of $X$ is high or when there are many treatments.

In conclusion, we propose inference methods for matching procedures to test the global null hypothesis of no difference between treatments, as well as estimate pairwise treatment effects with multiple treatments that are generally valid, accurate, and precise. These inference methods can be easily adjusted to estimate average differences from a control and the overall mean.

\section*{Aknowledgement}

This research was supported through a Patient-Centered Outcomes Research Institute (PCORI) Award ME-1403-12104. Disclaimer: All statements in this report, including its findings and conclusions, are solely those of the authors and do not necessarily represent the views of the PCORI, its Board of Governors or Methodology Committee. This research was also supported by the National Institute of Arthritis and Musculoskeletal and Skin Diseases of the National Institutes of Health (NIH) under Award number R01AR056328.

\section*{Financial disclosure}

None reported.

\section*{Conflict of interest}

The authors declare no potential conflict of interests.

\newpage

\bibliographystyle{wileyj}
\bibliography{references}

\end{document}